# Flow-pattern switching in a Motored Spark Ignition Engine

**Preeti S. Abraham[1], Xiaofeng Yang[2], Saurabh Gupta[1,2], Tang-Wei Kuo[2], David L. Reuss[1], Volker Sick[1]**

[1] Department of Mechanical Engineering,
The University of Michigan, Ann Arbor, MI, USA

[2] Propulsion Systems Research Laboratory,
General Motors Global Research & Development, Warren, MI, USA

**Corresponding Author:**

David L. Reuss, PhD., Research Scientist, Department of Mechanical Engineering, University of Michigan,1012 Lay Automotive Lab,1231 Beal Ave.,Ann Arbor, MI 48109-2133
Email: dreuss@umich.edu

**ABSTRACT**

Cyclic-to-cycle variability, CCV, of intake-jet flow in an optical engine was measured using particle image velocimetry (PIV), revealing the possibility of two different flow patterns. A phase-dependent proper orthogonal decomposition (POD) analysis showed that one or the other flow pattern would appear in the average flow, sampled from test to test or sub-sampled within a single test; each data set contained individual cycles showing one flow pattern or the other. Three-dimensional velocity data from a large-eddy simulation (LES) of the engine showed that the PIV plane was cutting through a region of high shear between the intake jet and another large flow structure. Rotating the measurement plane $\pm 10°$ revealed one or the other flow structure observed in the PIV measurements. Thus, it was hypothesized that cycle-to-cycle variations in the swirl ratio result in the two different flow patterns in the PIV plane.

Having an unambiguous metric to reveal large-scale flow CCV, causes for this variability were examined within the possible sources present in the available testing. In particular, variations in intake-port and cylinder pressure, lateral valve oscillations, and engine RPM were examined as potential causes for the cycle-to-cycle flow variations using the phase-dependent POD coefficients. No direct correlation was seen between the intake port pressure, or the pressure drop across the intake valve, and the in-cylinder flow pattern. A correlation was observed between dominant flow pattern and cycle-to-cycle variations in intake valve horizontal position. RPM values and in-cylinder flow patterns did not correlate directly. However, a shift in flow pattern was observed between early and late cycles in a 2900-cycle test after an approximately 5 rpm engine speed perturbation.

**INTRODUCTION**

Cycle-to-cycle flow variations in internal combustion engines are a leading cause of cycle-to-cycle combustion variations, such as misfires and partial burns, particularly in stratified-combustion mode operation. In this study, PIV measurements are made in a motored optical engine at a nominal steady state, but with small RPM and pressure boundary perturbations, in order to examine cycle-to-cycle flow variations without the added complexity of combustion or transient boundary conditions. Cycle-to-cycle flow variations are commonly understood to be variations in large-scale, high-velocity flow structures, while small-scale flow variations are held to be turbulence [1-3]. However, the traditional Reynolds Averaged Navier-Stokes (RANS) decomposition considers all flow variations about the mean flow to be turbulence. Thus, studies have used spatial filtering to separate fluctuations about the mean flow into large-scale cycle-to-cycle flow variations and small-scale turbulence [4-6]. The size of the spatial filters, i.e. the spatial cut-off frequency, was determined using various metrics; these included the temporal autocorrelation function [4], the turbulence power density spectrum [5], the fast-Fourier transforms of the ensemble average velocity [4, 6] or the total fluctuating velocity component [7], and the calculated motion of the flow structures [3]. Attempts have also been made to analyze cycle-to-cycle flow variability using tumble or swirl ratios [7-10]. Tumble or swirl ratios are measures of the dominance of the large-scale tumble or swirl vortex in the in-cylinder flow. Thus, examining the mean and variance of tumble or swirl ratios in a highly directed flow provides an estimate of the cycle-to-cycle flow variation of the largest length scale of the flow.

In this paper, proper orthogonal decomposition (POD) was used to quantify the cycle-to-cycle flow variations within individual data sets and evaluate potential causes. POD has been used in previous studies to differentiate between cycle-to-cycle flow variations and turbulence by choosing a cut-off mode, with lower order modes being used to reconstruct the cycle-to-cycle flow variations and higher order modes used to reconstruct the turbulence [1, 11]. However, Cosadia et al.[12] found that determining a cut-off mode was not possible, as the cumulative energy fractions of the POD of the instantaneous velocity fields converged very slowly despite the presence of a strong in-cylinder swirl motion. Enaux et al. [13] used the ratio between the energy fractions of the first and second modes as a measure of cyclic flow variations. Chen et al. [14, 15] proposed that, if the first mode is a good estimate of the mean flow, the POD coefficients of the first mode may be used to estimate the mass-specific kinetic energy contribution of the mean flow to any velocity field used in the decomposition, thus estimating how closely a particular velocity field compares to the mean flow. Similarly, the mass-specific kinetic energy contribution of all higher modes estimates the RANS turbulence in a specific velocity field. Thus, the cyclic variability of the mean flow and of the turbulence may be quantified. These studies performed the POD analysis on velocity fields at single crank angles from several cycles (phase-dependent POD). In this paper, POD was used to identify cycle-to-cycle variations between data sets in the large-scale flow structures during the intake stroke. In addition, local spatial averaging of velocity is also presented as an alternate method for quantifying cycle-to-cycle flow variations within a data set. However, unlike POD, prior identification of cycle-to-cycle variations in flow patterns was required to effectively choose an area in which the velocity is averaged.

Three-dimensional large-eddy simulation (LES) flow results are used to better understand the cycle-to-cycle flow variations seen in the central vertical plane of the engine in both experimental and computational data. LES is a tool for advanced IC engine design, explicitly capturing the dynamics of the large scales. A key advantage of LES when applied to IC engines is its ability to model cycle-to-cycle variations, unlike RANS [16-19]. Relevant studies by other researchers include examinations of turbulence statistics for a motored engine configuration [20, 21] and validation of cycle-to-cycle variations for a motored single-cylinder piston engine using LES [13]. Previous studies have also shown that LES computations for engines agree well with experimental results [13, 22, 23], including an earlier configurations of the engine presented in this paper [16, 24]. In addition, Schiffmann et al. [25] have continued the experimental efforts to guide the evaluation of LES model selection for the third generation of this engine as part of a larger study of stochastic flow in internal combustion engines [26-28].

However, the observed flow CCV and the occurrence of two distinctly different intake-jet flow patterns at 100° ATDCE provide an opportunity to explore potential causes of the large scale CCV; in particular, intake-port and in-cylinder pressure variations, unintended engine-speed variations, and unintended intake valve oscillations provided this opportunity. The manifold flow dynamics are considered to be one possible source of CCV [29, 30], which motivated the look at the intake-port pressure. Intake-port pressure oscillations, in-cylinder pressure and engine speed are all coupled as the piston speed drives the pressure drop across the intake valve that further affect the mass flow rate of air into the cylinder during the intake stroke. This motivated the investigation here into engine speed (rpm) perturbations that were caused by a dynamometer malfunction during one of the 3000-cycle (8 min) tests. The lateral intake-valve oscillations or 'valve ringing', where the valve oscillates like a pendulum during opening and closing, was due to a combination of long valve stems (145 mm) in this engine and valve-guide wear, a phenomenon also exhibited by older production engines. Changes in large-scale in-cylinder flow structures with changing intake-valve configurations have been observed in motored engines by other researchers [31-33], and one would logically expect the lateral oscillations would affect the azimuthal mass-flow distribution. Thus, the cycle-to-cycle variations in intake valve position observed here provided an opportunity for study.

In the following section, the optical engine and high-speed particle image velocimetry (PIV) experimental setup are presented, followed by a brief description of the LES method, which was used for a three-dimensional interpretation of the two-component PIV data. The results begin by showing the test-to-test variability and intra-cycle evolution of the ensemble average flow, as well as CCV. Next, phase-dependent POD is used to quantify the existence of one or the other distinctive patterns observed at 100° ATDCE, revealing test-to-test, intra-test, and cycle-to-cycle variability. The LES data is used to demonstrate CCV in the swirl ratio can result in the two observed flow patterns. Finally, the POD mode coefficients related to this flow pattern are used, as quantitative metrics, in an attempt to quantify the potential causes for the cycle-to-cycle flow variations; namely, intake-port and cylinder pressure variations, RPM variations, and lateral intake-valve oscillations.

**EXPERIMENTAL SETUP AND DATA ACQUISITION**

The optical engine used to collect the PIV data presented in this paper, TCC-I, is a two-valve single-cylinder pancake-chamber engine with a simple geometry that was designed for both high optical accessibility and for easy creation of a computational mesh. The original TCC-0 engine [34-36] has been modified for these tests with new intake and exhaust systems, a full quartz cylinder, and valve-seats reground from a one-angle to a two-angle seat profile. Although the present installation of this engine has only been run motored, a spark plug was included in the build so that the in-cylinder flow in its vicinity would be more realistic. The ground strap of the spark plug points towards the intake valve and is in the plane that bisects the valves. The engine specifications are given in Table 1. Crank angle degrees are specified as after top dead center exhaust (ATDCE).

| | |
|---|---|
| Bore | 92 mm |
| Stroke | 86 mm |
| Connecting Rod Length | 231 mm |
| Crank Radius | 43 mm |
| Geometric Compression Ratio | 10:1 |
| Displacement | 0.57 L |
| Clearance Volume | 0.064 L |
| Intake Valve Opening | 712° ATDCE |
| Intake Valve Peak Lift | 114° ATDCE |
| Intake Valve Closing | 240° ATDCE |
| Exhaust Valve Opening | 484° ATDCE |
| Exhaust Valve Peak Lift | 606° ATDCE |
| Exhaust Valve Closing | 12° ATDCE |

*Table 1: TCC-II engine geometry and valve timings*

Pressure data was measured every 0.5 crank angle (CA) degree at the intake plenum inlet, the exhaust plenum outlet, the interface between the intake and exhaust runners/ports, and within the cylinder. The intake and exhaust pressure transducers had a precision of 0.03 kPa based on one standard deviation. The cylinder pressure during the open part of the cycle was measured with a precision of 0.12 kPa. The data presented in this paper are four 66-cycle samples from three separate tests operated with nominal set points of 800 rpm, $P_{IntkPlenIn}$ = 95 kPaA, and $P_{ExhPlenOut}$ = 101.5 kPaA. Data Sets #2 and #3 had an RPM standard deviation less than 1 rpm and plenum and port pressure standard deviation less than the 0.03 kPa precision. However, in Data Set #1, acquired over approximately 7 minutes, a dynamometer controller malfunction led to RPM transients of ± 8 rpm and $P_{IntkPlenIn}$ transients of ± 0.8 kPa, which will be detailed later.

The high-speed PIV measurements were made using a dual-cavity diode-pumped neodymium-doped yttrium lithium fluoride (Nd:YLF) laser (Quantronix, Darwin-Duo). The laser sheet created had a thickness of 2 mm and was placed in the tumble plane bisecting the valves and containing the spark plug ground strap. Various high-speed complementary metal oxide semiconductor (CMOS) cameras were used to acquire the different data sets (Vision Research, Phantom v7.1, v7.3, and v1610). A 210 mm focal length camera lenses (Nikon

Micro-Nikkor 210 mm ED) was used for imaging. The intake air flow was seeded with silicone oil droplets with a nominal diameter of 1 µm. The raw PIV images were acquired and processed using a commercial software package (LaVision, DaVis Version 7). Details of the PIV processing, image correction for the cylinder wall, and the error analysis may be found in Abraham et al. [37]. The resulting vector fields from all data sets were mapped onto a common grid, depending on initial magnification and camera model used, the original spatial resolution was between 2.93 mm with a separation between velocity vectors of 1.46 mm and 1.59 mm with a separation between velocity vectors of 0.80 mm for the high-resolution data.

Figure 1 shows the experimental setup. The field-of-view of the camera (refer to Figure 2) included the tumble plane from the piston at bottom dead center (BDC) to the engine head.

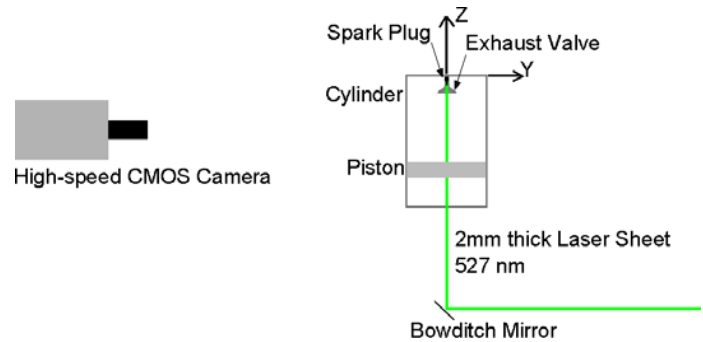

*Figure 1: Experimental setup*

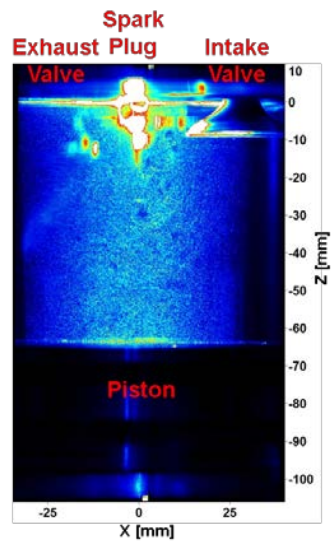

*Figure 2: Calibrated PIV image*

*Data Set #1* contains 2900 PIV image pairs acquired during consecutive cycles at 100° ATDCE and 800 rpm engine speed. PIV data from two 66-cycle subsamples, SS1 (cycles 201-266) and SS2 (cycles 2501-2566), are investigated in this paper. The magnification was 0.180 along the X axis and 0.171 along the Z axis. The time difference (dt) between the PIV image pairs was 20 μs, resulting in a velocity dynamic range of approximately 0.9 m/s to 50 m/s. This data set was acquired early in the program and rejected for the purpose of LES validation because of both the static dt available at that time and a dyno-controller malfunction. However, it was resurrected here specifically because the flow patterns switched one to the other after the 700 cycle, 8 rpm (1%) engine-speed perturbation. Although the larger velocities in the jet are not resolved, the lower velocities in the region of the pattern studied here are resolved. *Data Set #2* contains PIV image pairs acquired during 66 consecutive cycles from 0° through 360° ATDCE, every 2°, at 800 rpm. An adaptive PIV setup was set up to change dt throughout the cycle [37] that was used for Data Sets #2 (and #3) to resolve the larger intake-jet velocities. The dt was 10 μs during intake of Data Set #2, resulting in a velocity dynamic range of 2 m/s to 108 m/s that better resolved large intake-jet velocities. The magnification was 0.163. *Data Set #3* contain PIV image pairs acquired during 66 consecutive cycles from 0° through 715° ATDCE, every 5°, at 800 rpm. A time delay of 6 μs during intake, resulted in a velocity dynamic range of 2 m/s to 122 m/s. The magnification was 0.306. Data Sets #2 and #3 were chosen because their 66-cycle ensemble-average velocity from each test showed either one or the other patterns, respectively.

**LES ENGINE MODELING**

Three-dimensional, compressible, unsteady and turbulent cold flow large-eddy simulations for the optical engine presented in this paper were carried out at General Motors Research & Development using CONVERGE [38], a commercial computational fluid dynamics code. The wave dynamics of the complete engine experimental setup was simulated using GT-Power, a detailed 1-D flow model. The initial pressure and temperature at intake plenum/port, cylinder, and exhaust port/plenum are based on the 1-D result at TDC. Initial turbulent kinetic energy and its dissipation are set to 1 $m^2/s^2$ and 100 $m^2/s^3$ respectively. The total pressure calculated using GT-Power and temperature were applied at in-flow boundaries at the inlet of the plenum to drive all simulations. The static pressure is set as boundary condition at the exit of the exhaust plenum. A one-equation eddy viscosity model [39-42] was used to model the turbulent viscosity. The Werner-Wengle wall model [43] was enabled close to the wall boundaries where the smallest mesh dimension is set to 0.5 – 1.0 mm, corresponding to a dimensionless wall distance $y^+ > 11.2$. In order to capture the flow structures accurately, a second-order numerical differencing scheme was used for the convective terms in the momentum equation. The PISO algorithm [44] was used for pressure-velocity coupling which resulted in a temporal accuracy comparable to a second-order scheme. The sub-grid stress tensor modeled by LES has second-order accuracy.

The mesh arrangement is shown in Figure 3. In order to minimize the effect of mesh topology on the computed results, a special feature of the code called 'embed sphere' is employed with the finest mesh size of 0.5 mm used in valve seat and spark plug regions and the coarsest mesh size of 8 mm placed inside the intake and exhaust plenums. Orthogonal cubic meshes are automatically created at run time to avoid any variation in cell shape or size.

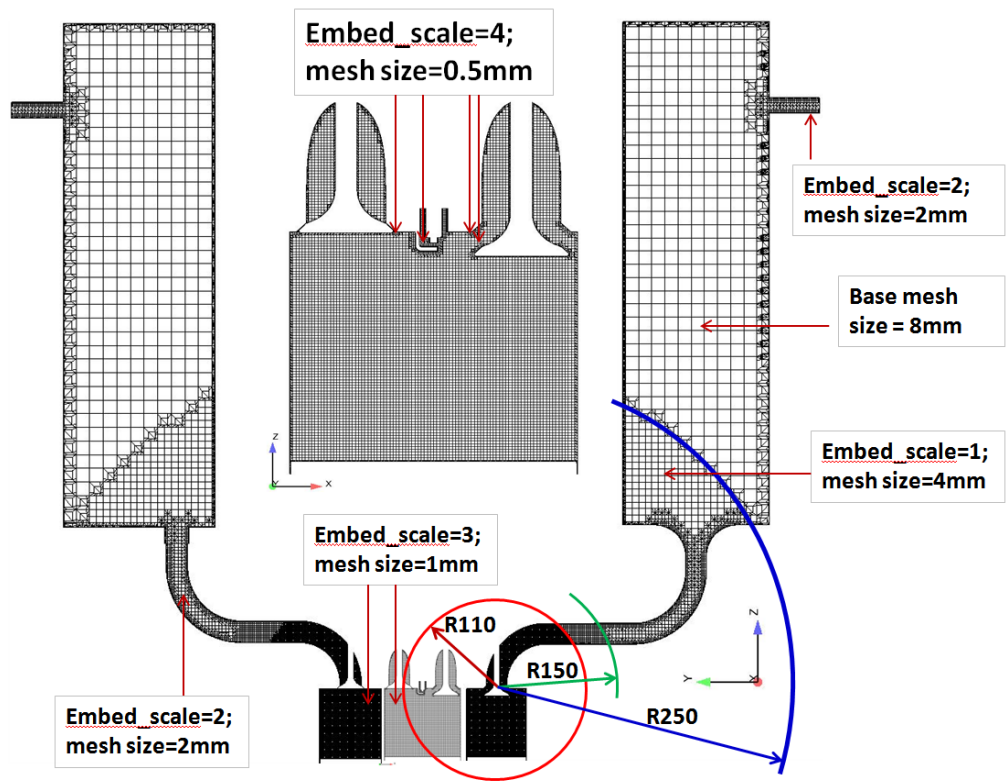

*Figure 3: Mesh arrangement with in-cylinder mesh size of 1 mm. Note that the enlarge insert shows the cylinder cross section in which experimental data were obtained. The left and right cross sections, which include the plenums and runners, are normal to the measurement plane and are shown rotated here to illustrate the shape of the intake system.*

A hardware upgrade on the engine head resulted in small changes in the valve seat geometry from the original TCC-I 1-angle design used in the LES computations to the TCC-II 2-angle valve seat geometry that existed in the engine, as shown in Figure 4. Incorporating these changes into the computational mesh may result in some changes in the in-cylinder flow. However, both valve seat designs were found to produce similar in-cylinder flow patterns, with a large-scale high-shear region between the intake jet and the large-scale entrainment vortices.

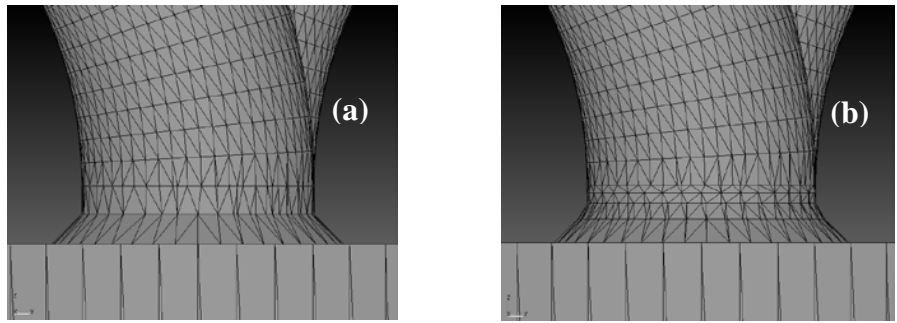

Figure 4: (a) 1-angle; and (b) 2-angle valve seat geometries

In-plane 2-D velocity components for various cutting planes are extracted from the three-dimensional LES computed data for individual cycles, and then averaged over 56 cycles to obtain the mean velocity contours. All LES simulations were conducted at a fixed engine speed of 800 rpm without lateral valve oscillations. However, both engine speed and valve stem oscillations are observed in experiments and their effects on PIV measurements will be discussed in the following sections.

**RESULTS**

*In-cylinder flow variability at 100° ATDCE*

The two different ensemble average flow patterns were initially observed during a comprehensive investigation at mid intake stroke (100° ATDCE). The two different flow patterns are seen in the ensemble average velocity fields from two different 66-cycle samples from one 2900-cycle test ( #1, SS 1 and 2), and two other shorter tests (#2, and #3), shown in Figure 5. The flow past the open intake valve (see Fig. 2 for location details) produces a strong jet of air in the vertical PIV cross section. This is seen near the centerline in the images from the top downward. The images shown here and analyzed further have been cropped at – 20mm to remove the impact of restricted dynamic range in data set #1 from the following discussion. In addition we highlight the flow structure in the lower-right-hand (LRH) quadrant of the field-of-view with a blue box and an arrow indicating the major flow direction in this region. Similarities and differences between the different ensemble average velocity fields in Figure 5 are quantified using relevance indices. The relevance index, $RI_{U,V}$, is computed by projecting one velocity field (U) onto another (V) and is defined as shown in Equation (1).

$$RI_{U,V} = \frac{(U,V)}{\|U\|.\|V\|} \tag{1}$$

In Equation (1), the numerator is the inner product of two velocity fields over the whole domain. If the normalized flow patterns in U and V are identical, $RI_{U,V}$ will be equal to 1. U and V are orthogonal if $RI_{U,V}$ is equal to 0 [45].

Figure 5 shows that the ensemble average velocity fields from data sets #1, SS1 and #2 are similar (as indicated by the relevance index above 0.9). The ensemble average velocity fields from data sets #1, SS2 and #3 are also similar to each other, with upward flow toward the upper-left-hand corner. Nonetheless, the RI between #1, SS 1 and #1, SS2, and the RI between #2 and #3 are only reasonably high because the flow structures in the other three quadrants are similar. Though the 66-cycle samples should give reasonable estimates of the ensemble average it is clear that there are different flow characteristics between the samples. *The key point illustrated by Data Set #1 is that the occurrence of the different flow patterns was not due to test-to-test operation variability but can occur during the course of a single test.* The occurrence of two states of this LRH flow structure raised three questions: (1) How does the flow change during the cycle, (2) how does the state change from cycle to cycle, and (3) what caused the measured flow to change from one state to the other?

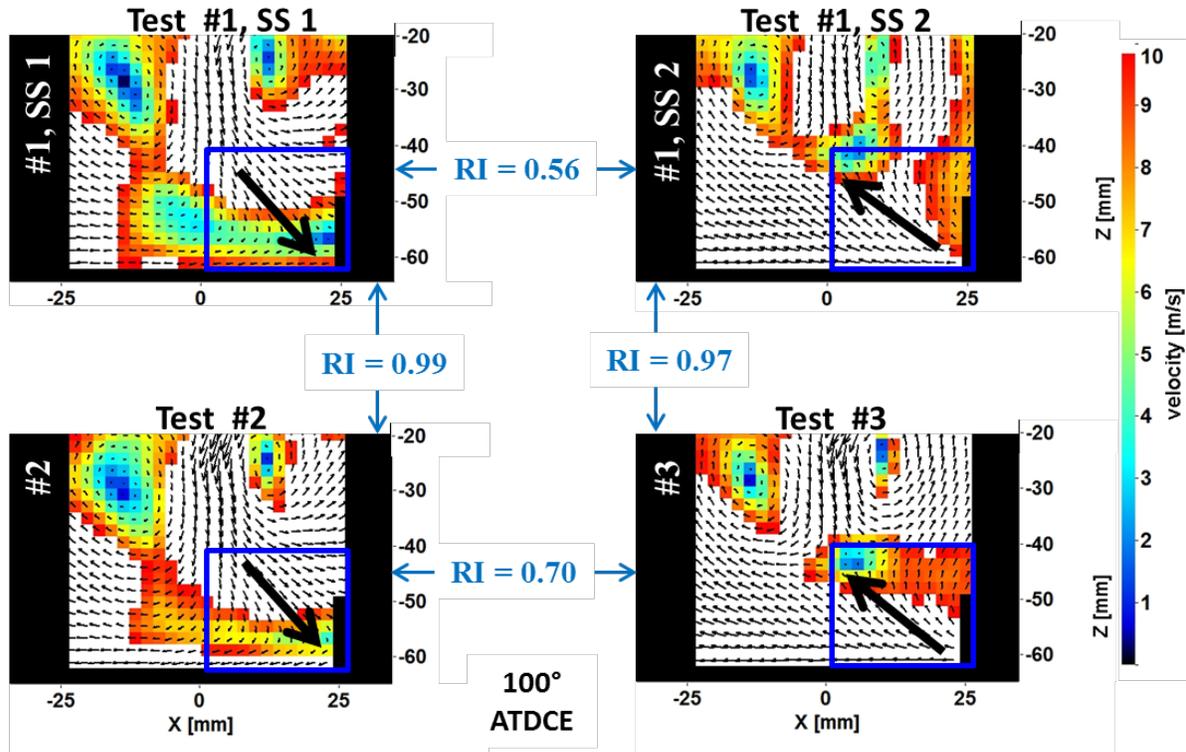

*Figure 5: Ensemble average velocity fields from data sets #1, #2, and #. The Relevance Index RI is used to quantify the equivalency (or lack of equivalency) between the flow patterns.*

*Intra-cycle flow variability* was investigated to determine if the flow pattern difference is unique to the intake jet or persistent throughout the cycle. Figure 6 shows the ensemble-averaged flow patterns from Data-Sets #2 & #3, between TDC Intake and 240 ATDC intake valve closing, IVC. The circles highlight regions for comparison between the two tests (left to right). The key point of this figure is that the test-to-test difference is persistent throughout the intake cycle, The exception is 140 ATDC, where no obvious differences are observed; this was included to illustrate the importance investigating all crank angles. Figure 6 demonstrates that the two distinctively different patterns observed at 100 ATDCE are indicative of a dual flow-state through the cycle. For pragmatism, the flow at 100 ATDCE will be used as a metric for the two flow states and the focus of the remainder of the paper.

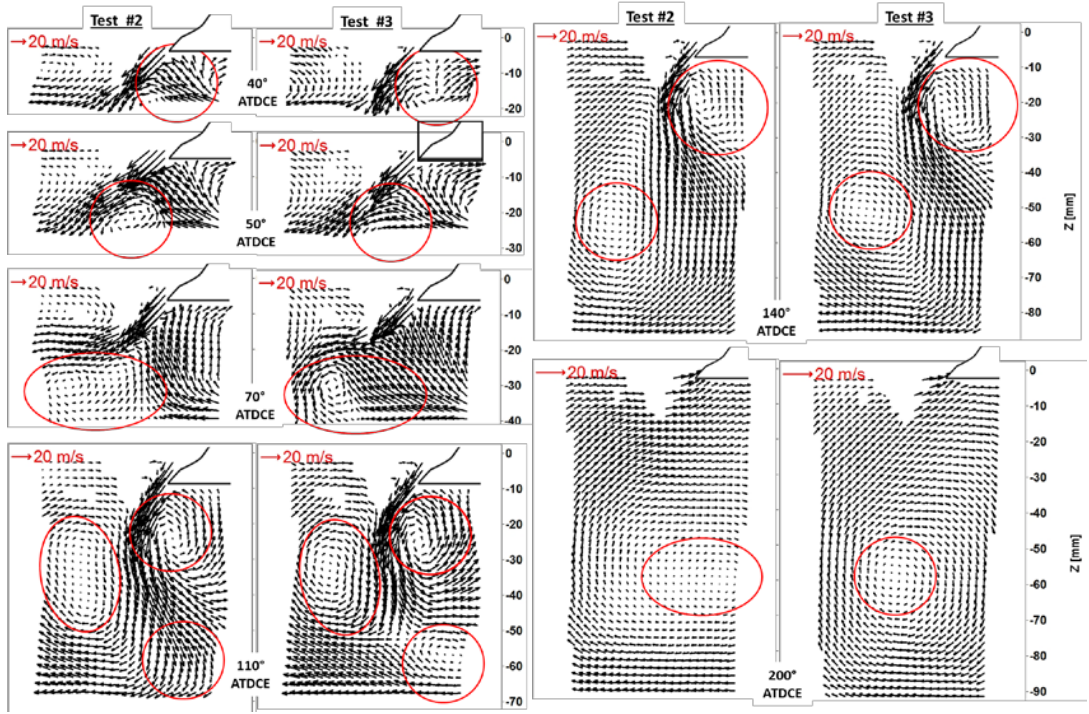

*Figure 6  Ensemble-averaged velocity distributions from Data Sets #2 and #3, between TDC exhaust and intake valve opening. Circles highlight regions of comparison.*

To illustrate *CCV of the two flow patterns*, Fig. 7 shows two consecutive cycles (36 & 37) from Data Set #3. It can be seen that the clearly downward (negative w) velocity component exists in Cycle 37 for extended time. Figure 7 illustrates that the occurrence of these two flow pattern states is not a long-time scale variability but rather CCV, statistically more or less probable depending on the samples used for the ensemble averaging.

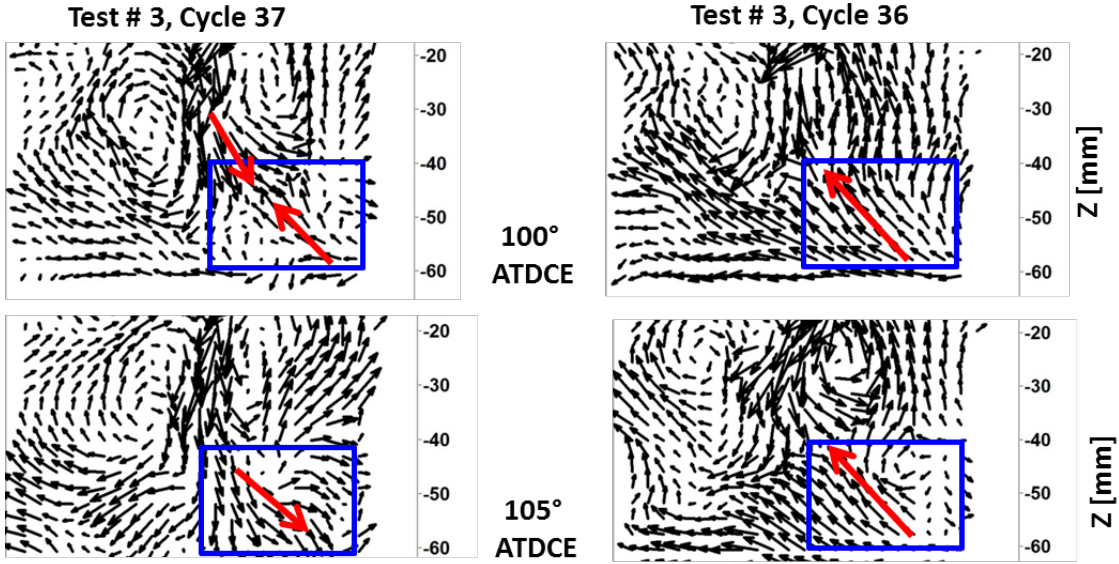

*Figure 7 Illustration of the instantaneous flow pattern CCV between two consecutive cycles in Data Set #3, near 100 ATDCE.*

Phase-dependent POD analysis was utilized to quantify the CCV of these two flow states in the direction of this flow structure in each cycle at 100 ATDCE. Using phase-dependent POD, the velocity fields from the combined four data sets were analyzed to obtain a common set of basis functions or modes [46, 47]. These modes are ordered by energy fraction, which is the average mass-specific kinetic energy percentage contributed by a mode to the velocity fields analyzed. Figure 8 shows the first two modes from the phase-dependent POD analysis, with energy fractions of 55% and 14%, respectively, thus together capturing 69% of the total energy. Higher order modes have energy fractions of 2% or less.

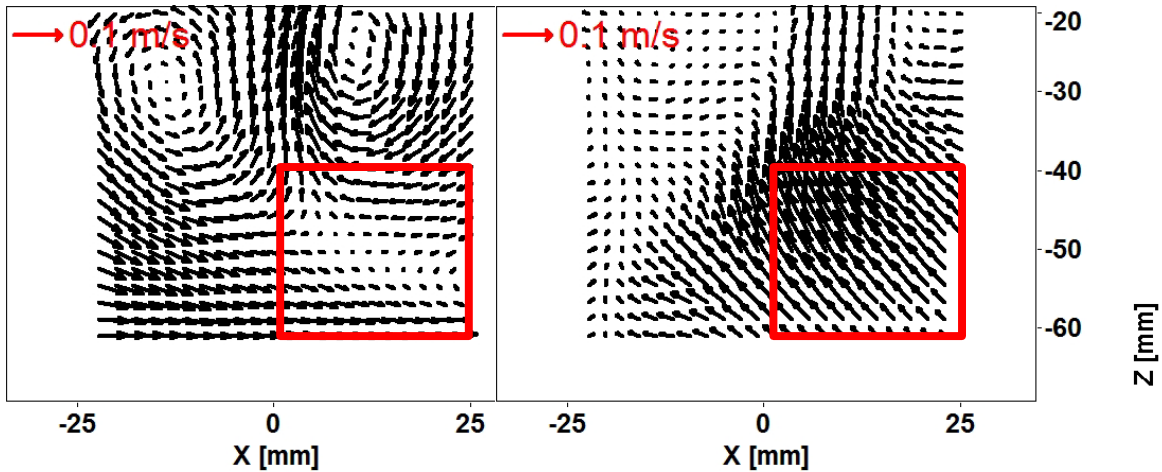

*Figure 8: Mode 1 (left) and mode 2 (right) from joint phase-dependent POD analysis of combined velocity data from data sets #1, SS 1 and 2, #2, and #3*

In Figure 8, Mode 1 shows the intake jet flow and the associated counter-rotating entrainment vortices, and Mode 2 shows the flow in the bottom-right corner of the field-of-view. Due to the sign ambiguity of POD modes, Mode 1 shows velocity vectors with signs opposite that seen in the velocity fields analyzed. When multiplied by the negative Mode 1 coefficients (shown in Figure 9), the contribution of Mode 1 to individual velocity fields estimates intake jet velocity distribution in each cycle (refer to [14, 15]). Similarly, Mode 2 estimates the state of the LRH flow structure in each cycle.

Figure 9 shows the mode coefficients associated with the first two phase-dependent POD modes computed for the combined four data sets. The Mode 1 coefficients are always negative because the jet that dominates Mode 1 is always in the same direction. However, the Mode 2 coefficients indicate the state of the LRH flow structure, with positive coefficients for cycles in which the flow in the bottom-right corner of the field-of-view is upwards and negative coefficients for cycles with downward flow in the bottom-right corner. Most of the Mode 2 coefficients for data sets #1, SS 1 and #2 are negative, indicating that most cycles in these data sets have downward flow in the bottom-right corner. Most of the Mode 2 coefficients for data sets #1, SS 2 and #3 are positive, indicating that most cycles in these data sets have upward flow in the bottom-right corner. This is consistent with the ensemble average velocity fields. However, Figure 9 also clearly shows that one velocity field in Data Set #1, SS 1 and three velocity fields in Data Set #2 have positive Mode 2 coefficients, illustrating cycle-to-cycle intake flow variations in these data sets. Similarly, three velocity fields in Data Set #1, SS 2 and several velocity fields from Data Set #3 have negative coefficients. This demonstrates that the CCV illustrated in Fig. 7 can be generalized, e.g., cycles can contain either state of the LRH flow pattern in a given data set, but one or the other tends to dominate.

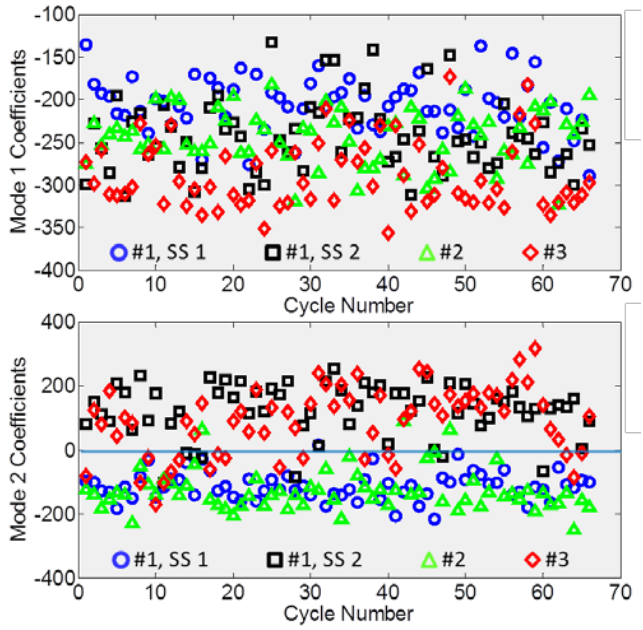

*Figure 9: Coefficients associated with the first two modes of the phase-dependent POD analysis of data sets #1, #2, and #3*

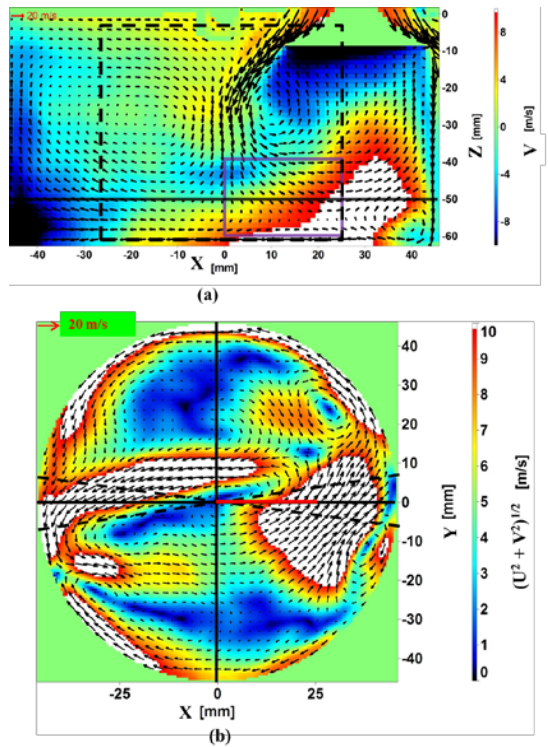

*Figure 10: Ensemble average three-component velocity from LES computations in (a) the Y = 0 plane (V > 0 is into the page; every second vector shown) and (b) the Z = -50 mm plane (Z > 0 is out of the page; every fourth vector shown). The dashed line box in (a) indicates the PIV field-of-view and the solid line box indicates the LRH region of interest in Figures 5 and 8.*

Investigation of the three-dimensional, LES data provided further understand of the CCV observed in the two-dimensional, Eulerian PIV-measurement plane. The 2-D Y = 0 and Z = -50 mm (black lines in Figure 10 (a)) planes from the 3-D simulation are shown in Figure 10. Inspection of the color contours in Figure 10 (a) reveals a large positive (into the page) out-of-plane velocity in the bottom-right corner of the cylinder (outlined in purple), which can also be observed in the horizontal plane (black line). Figure 10 (b) shows that the LRH region (red line) cuts across a region of high shear and flow reversal. Since the engine has a swirl ratio of about 0.9, it is reasonable to presume that this region could rotate through the Eulerian measurement place at Y = 0. This premise motivated investigating the change in the LRH region if the observation plane were rotated azimuthally by ± 10°. The results are shown in Figure 11 and demonstrate the same two states of the LRH region that were observed in the measurements. Thus, it is possible that cycle-to-cycle variations in the azimuthal location of this shear region through the measurement plane could explain the observed flow reversal in the LRH region. This is also in agreement with the observations made from Fig. 6 that shows the two flow patterns as the result of a phase shift from ~90-130 ATDCE in the evolution of the flow patterns. Independently performed phase-dependent POD (but not presented here for space reasons) shows such phase shifts in the mode coefficients of the mode that dominantly describes the LRH flow pattern. An analysis of in-cylinder tumble flow evolution with similar implications has been described by Voisine et al. [10].

Following the logic of the initial premise that the observed variations are the result of bulk swirl, the cycle-to-cycle variability of the swirl ratio was computed from the LES results. The swirl ratio, SR, is defined in Equation (2).

$$SR = \frac{\sum_{i=1}^{N} m_i \vec{r}_i \times \vec{u}_i}{\omega_c \sum_{i=1}^{N} m_i \vec{r}_i \cdot \vec{r}_i} \tag{2}$$

In Equation (2), $m_i$ and $u_i$ are the mass and velocity at a point i in the cylinder, $r_i$ is the distance from the center of the field-of-view to the point i, and N is the total number of points considered. $\omega_c$ is the engine speed measured in radians per second [8].

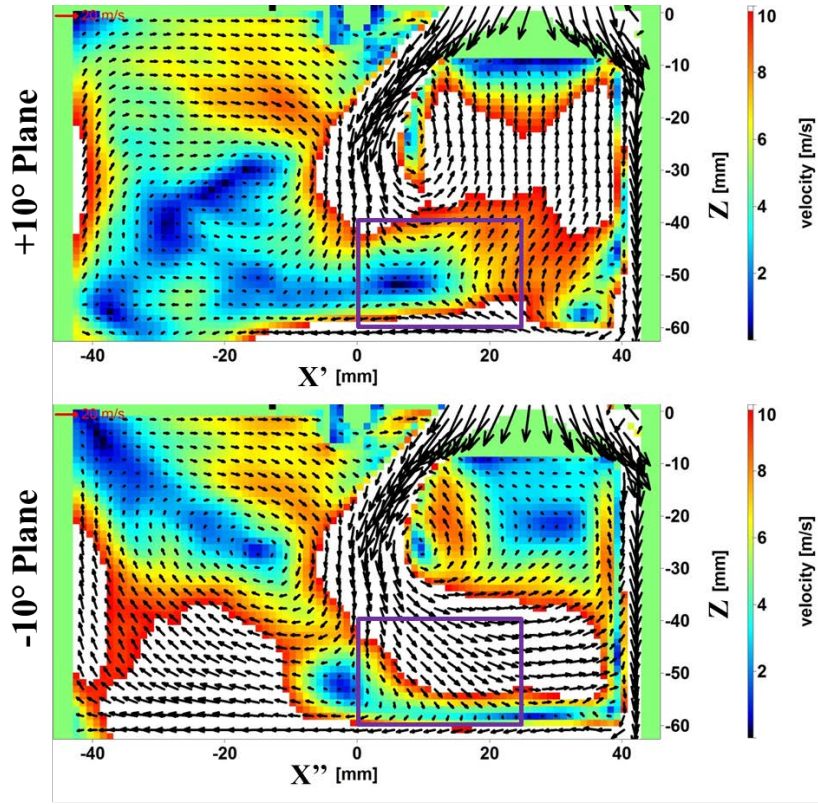

*Figure 11: Ensemble average in-plane velocity from LES computations in vertical planes ± 10° with respect to the Y = 0 plane (X'Z and X''Z, respectively; every second vector shown).*

Figure 12 indicates that LES predicts that the bulk motion, measured by tumble and swirl ratios, exhibits significant cyclic variations. Of particular interest here is the mass-averaged swirl ratio that shows a range of about 0.2 from cycle to cycle, with an average of about 0.9, which is a 22% variation. It is interesting to note that a solid body rotation would be expected to vary with a range of about 20° over the 100° duration of the intake stroke; of course, the assumption of solid body rotation is weak, but this just demonstrates it has the correct order of magnitude. It is as likely to expect spatial-phase variations of the large structures in Figure 9 from cycle to cycle. Regardless, the observations from the LES cycles demonstrate the potential hidden dangers of making comparisons with a stationary measurement plane.

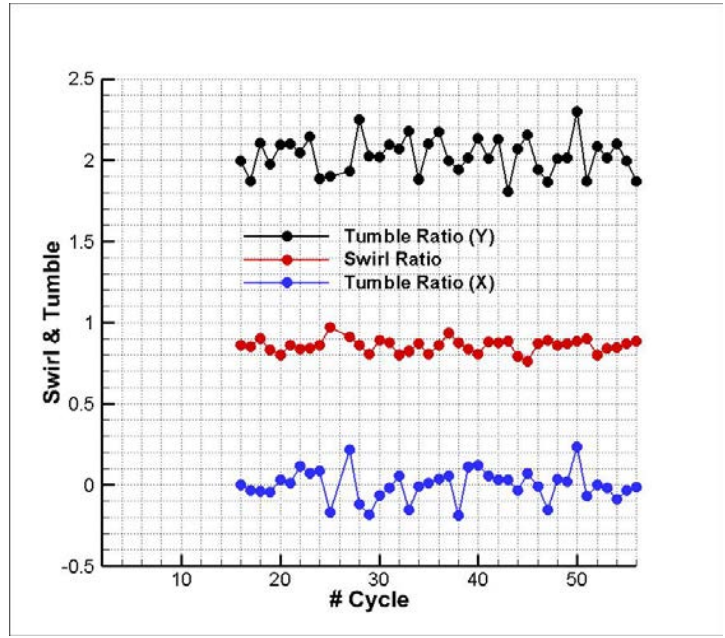

*Figure 12: Cycle-to-cycle variation in swirl and tumble ratios simulated with LES -potential causes of in-cylinder flow variability at 100° ATDCE.*

*Potential causes of the flow CCV*

Having established the POD coefficients as metrics for the occurrence of the two different states in the LRH region, and linked them to flow field cycle-to-cycle variability, it is of interest to seek the cause. Three potential causes of the observed in-cylinder flow variations at 100° ATDCE are examined in this section. First, cycle-to-cycle variations in the intake port and cylinder pressure boundary conditions might be expected to force changes in the mass flow through the intake valve and thereby affect in-cylinder flow patterns. Second, the intake valve was seen to experience lateral oscillations (like a ringing bell) during opening and closing. This geometric boundary change would change the annular symmetry of intake-valve curtain area and thus affect intake jet flows. Third, cycle-to-cycle variations in RPM would also translate to cycle-to-cycle variations in mass flow. All three have the potential to trigger in-cylinder flow structure or swirl ratio variations. It is an advantage to have two data sets where the ensemble-average flows reveal the two patterns, since averaging can then be used on these parameters to variations that are small compared to the measurement noise.

*Cycle-to-cycle variations in intake port and cylinder pressure.* Figure 13 quantifies the level of cycle-to-cycle variations seen in the crank-angle resolved intake port pressures from the data sets presented in this paper. The spread in intake port pressure values at a particular crank angle over the course of a data set is about 1% of the mean intake port pressure, compared to about 0.03% RMS precision.

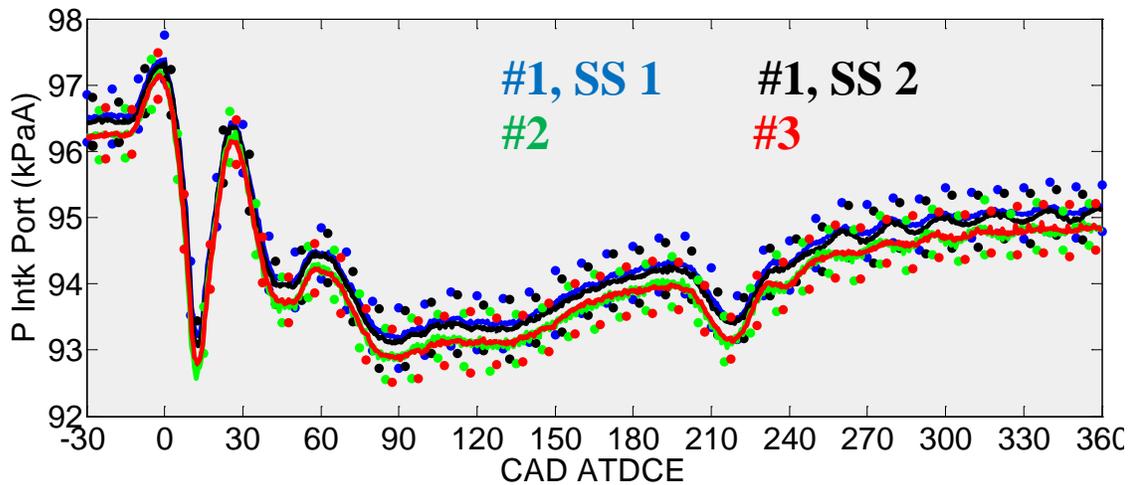

*Figure 13: Crank angle resolved intake port pressure data for all cycles from data sets #1, #2, and #3. The solid lines are the test ensemble-average and the points one standard deviation.*

Figure 14 examines the cycle-to-cycle variations in crank-angle resolved cylinder pressure from these data sets. The largest variations in cylinder pressure (a COV of approximately 0.2%) occur about TDC compression. The spike in the cylinder pressure standard deviation around 430 ATDC is before exhaust valve opening; it is speculated to occur due to small cycle-to-cycle cylinder-pressure variations during reversal of the piston ring from the bottom to the top of the ring groove.

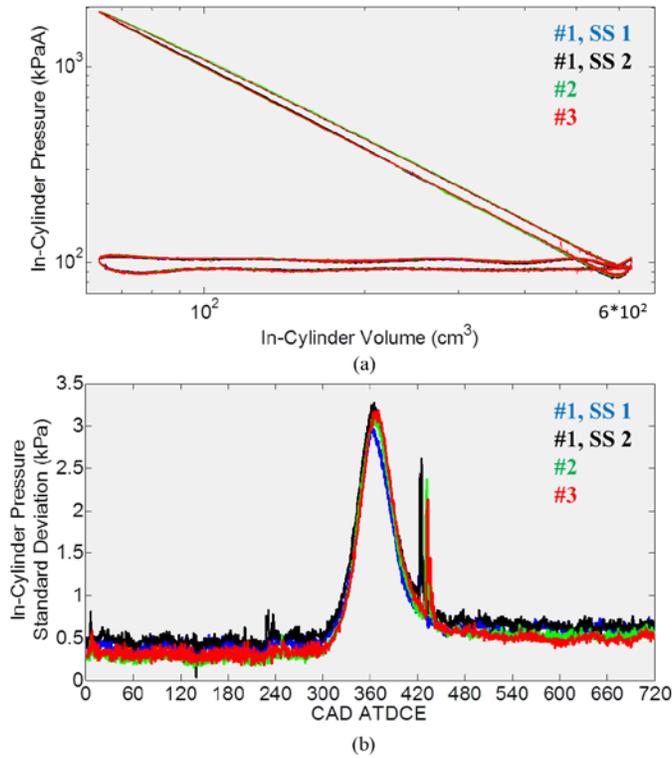

*Figure 14: (a) In-cylinder log P – log V graph for all cycles and (b) standard deviation for cylinder pressure in data sets #1, #2, and #3.*

To identify possible correlations, Figure 15 shows scatter plots of the 100° ATDCE intake port pressures and the intake valve pressure drop versus POD Mode 2 coefficients. Both the intake port and cylinder pressures were smoothed using running averages over 5° intervals to reduce digitization and transducer noise. The Mode 2 coefficients are, as shown earlier in this paper, indicators of the flow direction in the LRH region, and identical to the values in Figure 9. Figure 15 shows no indication of any systematic correlation, and, thus, these pressures can be ruled out as causes of the observed variability.

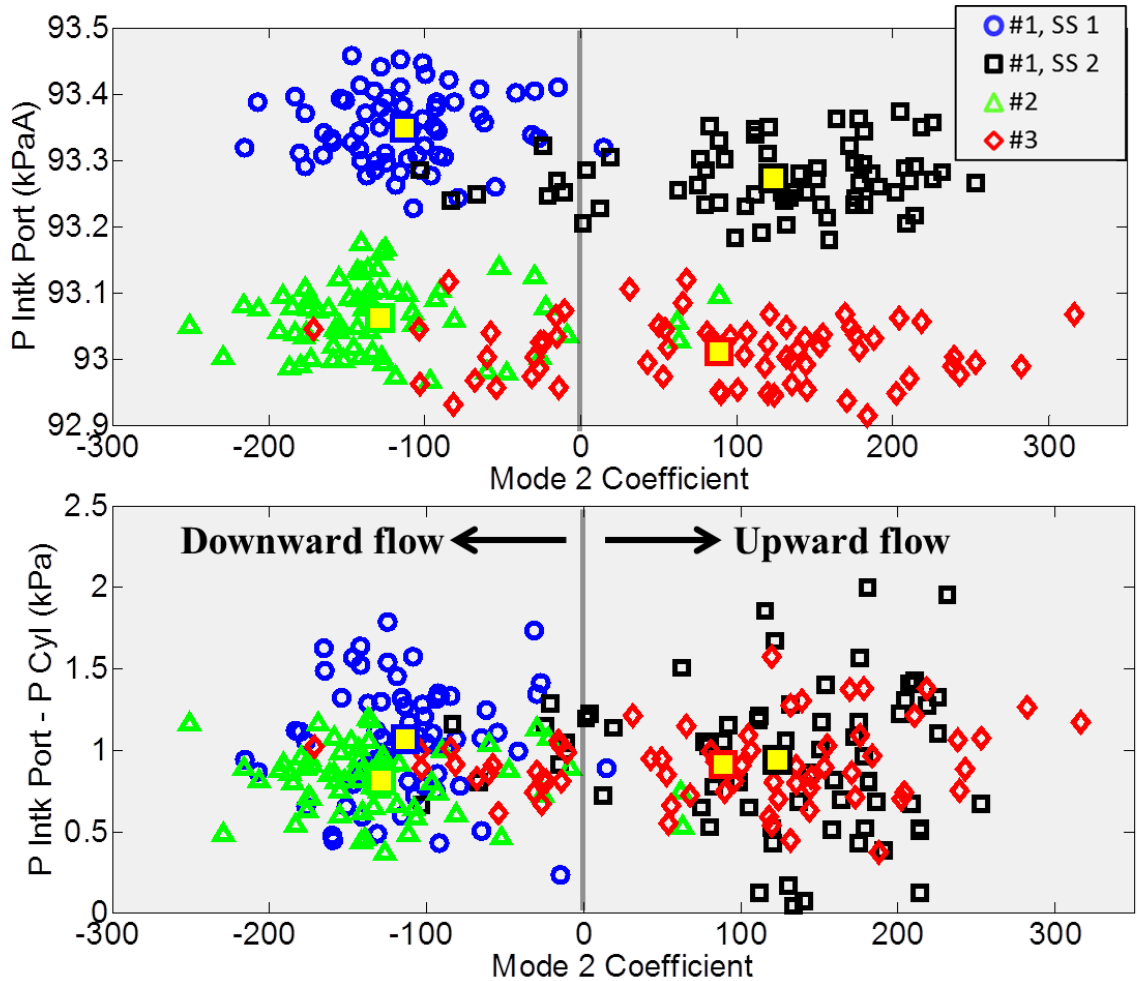

*Figure 15: Effect of intake port pressure and pressure drop across the intake valve on in-cylinder flow pattern at 100° ATDCE. The open symbols indicate values for individual cycles and the squares indicate ensemble average values.*

***Cycle-to-cycle variations in intake valve position.*** As a metric to correlate the valve oscillation with the flow direction, the variability in the lateral and axial (lift) location of the intake valve edges was measured at 100° ATDCE. The valve positions were measured from the calibrated PIV images (1 pixel = 50 μm), where they were illuminated by scattered laser light. The extreme locations of the intake valve edge were determined by calculating and comparing the maximum and minimum intensities at each pixel in the calibrated PIV images. The maximum intensity count of a particular pixel in the region of the valve image will be high if some part of the valve is ever imaged at that pixel. Conversely, the minimum intensity count of a particular pixel in the region of the valve image will be high only if some part of the valve is imaged at that pixel in every cycle. The valve locations were measured to the nearest pixel. An example of this process is shown in Figure 16.

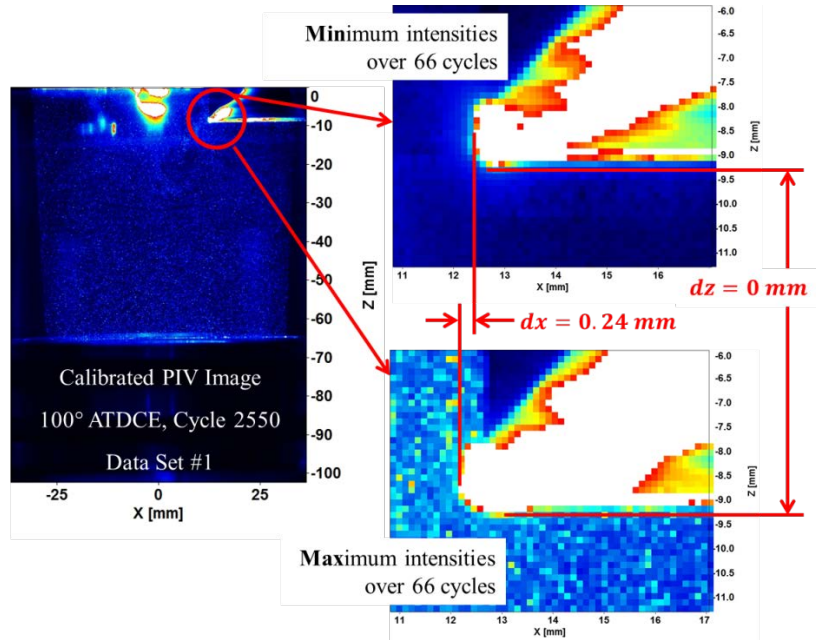

*Figure 16: Cycle-to-cycle variations in intake valve position at 100° ATDCE in Data Set #1, SS 2*

| Data Set | dx (mm) | dz (mm) |
|---|---:|---:|
| #1, SS 1 | 0.00 | 0.00 |
| #1, SS 2 | 0.24 | 0.00 |
| #2 | 0.00 | 0.14 |
| #3 | 0.15 | 0.10 |

*Table 2: Variability in intake valve position along X and Z axes*

The displacements along the X and Z axes between extreme valve positions at 100° ATDCE for all data sets presented in this paper are shown in Table 2. The cells in Table 2 that are not highlighted show displacements of 1 pixel or less. However, the highlighted cells in Table 2 show clearly that the intake valve position in data sets #1, SS 2 and #3 vary from cycle to cycle. These are the data sets dominated at 100° ATDCE by the flow pattern with upward flow from the bottom-right corner of the field-of-view. Of course, the valve is expected to have experienced out-of-plane (dy) motion not available in this view.

To observe correlation between intake valve position and in-cylinder flow pattern at 100° ATDCE, the Mode 2 coefficients of data sets #1, SS 2 and #3 were plotted against intake valve edge position (Figure 17). In Data Set #3, larger upward velocity vectors from the bottom-right corner are, on average, observed in cycles where the intake valve is closer to the centerline of the cylinder (X = 0). However, the Mode 2 coefficients for Data Set #1, SS 2 show no similar trend. This may partly be due to the lower spatial resolution of Data Set #1. Although these plots hint at a correlation, it is inconclusive with the single view and poor spatial resolution available here.

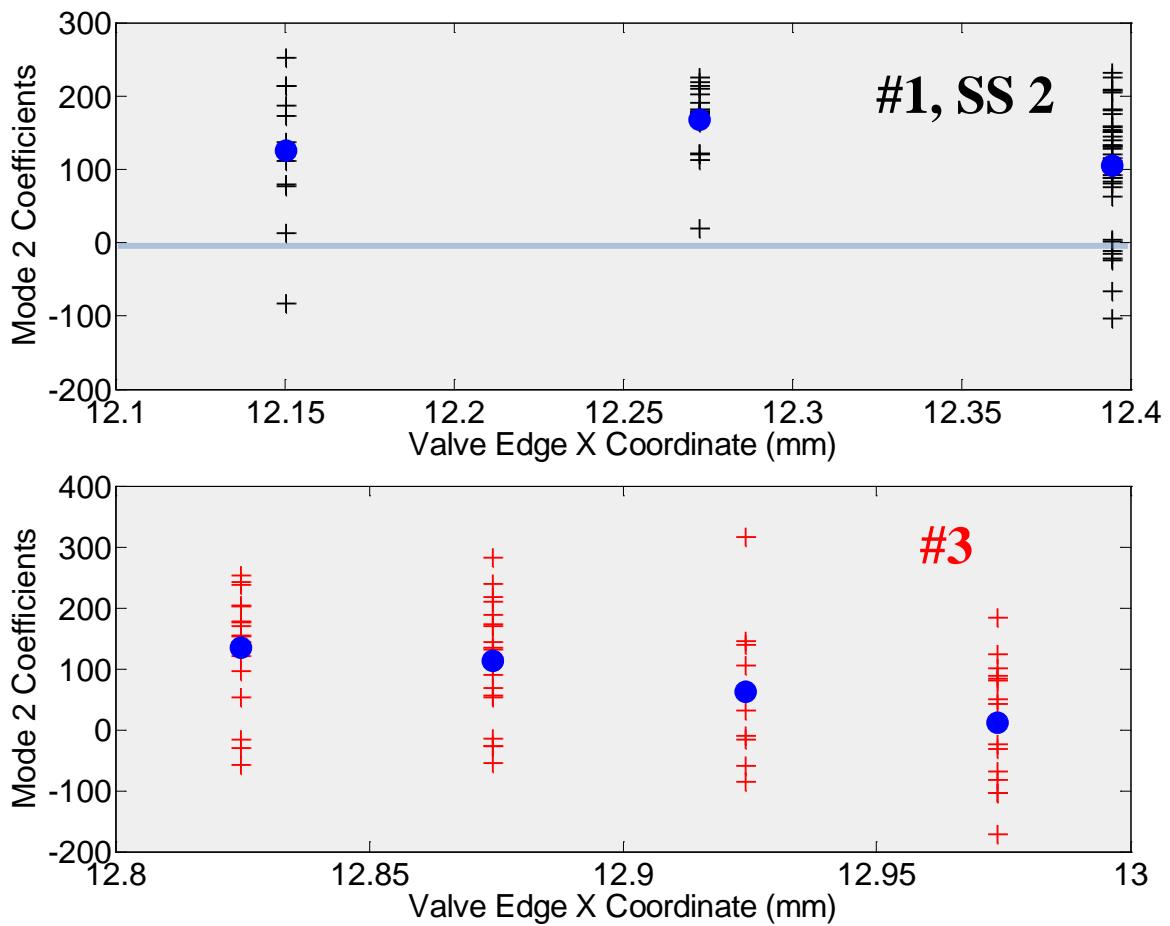

*Figure 17: Mode 2 coefficients (crosses) versus intake valve edge X coordinate. Average Mode 2 coefficient at each valve position is shown as a blue dot.*

*Variations in engine speed.* A random dynamometer controller malfunction lead to an eight rpm perturbation lasting approximately 800 of the 2900 cycles of test This data set provided a unique opportunity to study the potential impact of speed variation on in-cylinder flow. In order to examine correlations between engine speed and in-cylinder flow patterns, the cycle-to-cycle variations in RPM and the flow were compared for the 2900 cycles of Data Set #1, where the 66-cycle subsamples SS 1 and 2 showed the two different flow structures. The closed intake-air system provides a fixed air-mass flow rate, thus a change in engine speed will result in a change of intake pressure as well as CA-phase change of the intake-pressure waves. The cycle-to-cycle variations in RPM and the resulting variation in intake plenum inlet pressure are presented in Figure 18.

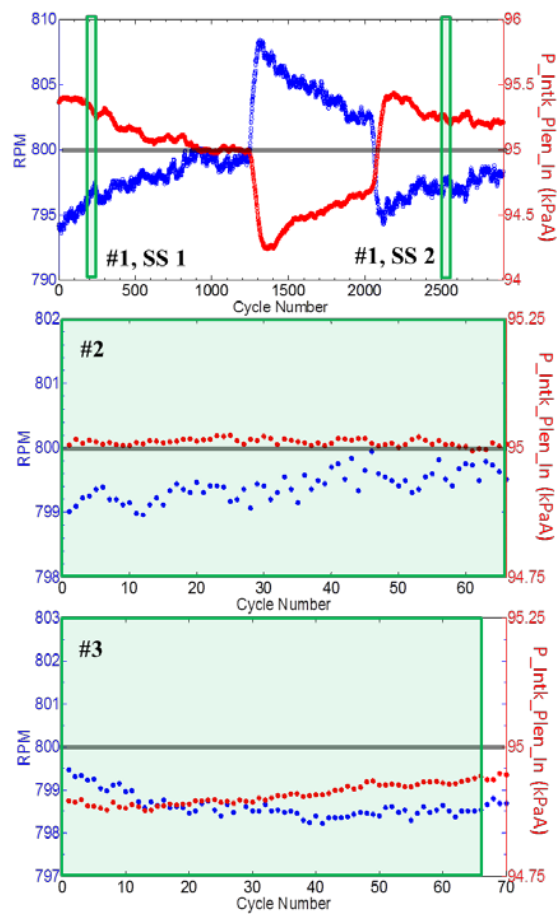

*Figure 18: Engine speed and intake plenum inlet pressure for data sets #1, #2, and #3. Engine operation set points are marked by grey lines and PIV cycles sampled are highlighted in green.*

Figure 18 shows that Data Set #1 experienced engine speed and, subsequently, intake pressure transients due to dynamometer operation errors. Figure 18 also shows that the RPM and intake plenum inlet average pressure data from data sets #2 and #3 vary by less than 0.25% from the desired set points.

The possibility of a one-to-one correlation between engine speed and intake flow patterns is considered by graphing RPM versus Mode 2 coefficients from the phase-dependent POD analysis at 100° ATDCE (Figure 19). However, no such correlation can be seen between RPM and intake flow pattern.

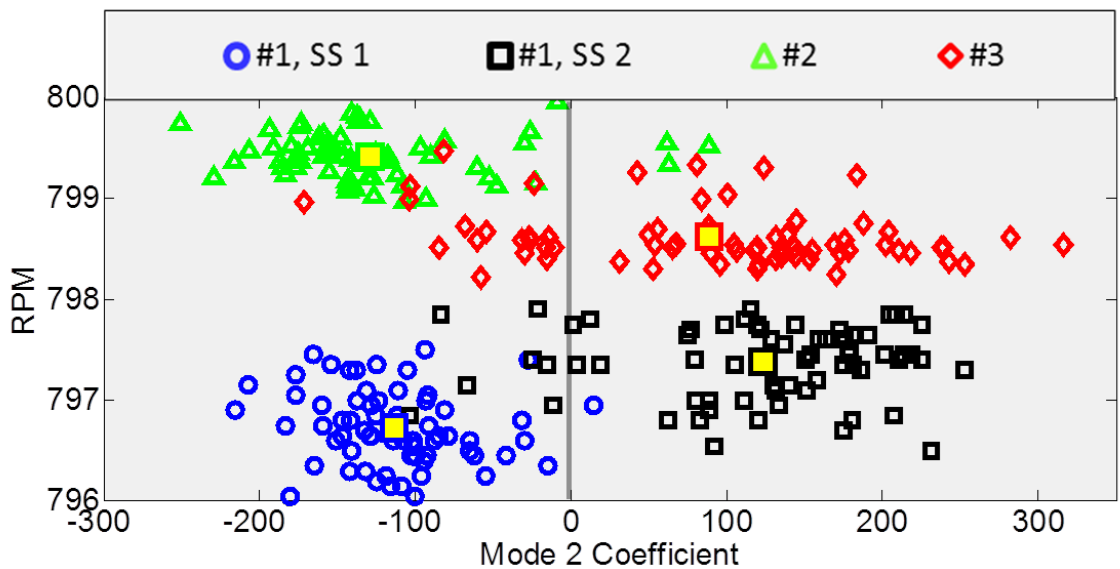

*Figure 19: Effect of engine speed on flow pattern at 100° ATDCE. The open symbols indicate values for individual cycles and the filled squares indicate ensemble average values.*

To quantify the flow pattern variation during Data Set #1, a spatial average of the velocity components was taken in a small area in the bottom-right corner of the velocity field-of-view, as shown in Figure 20. The use of the local average is a pragmatic alternative to the use of POD Mode 2, which would require the decomposition of 2900 cycles. The downward flow pattern (negative Mode 2 coefficient) will exhibit a positive local average velocity component along the X axis ($\bar{U}$) and a negative local average velocity component along the Z axis ($\bar{W}$), with opposite signs for the upward flow pattern.

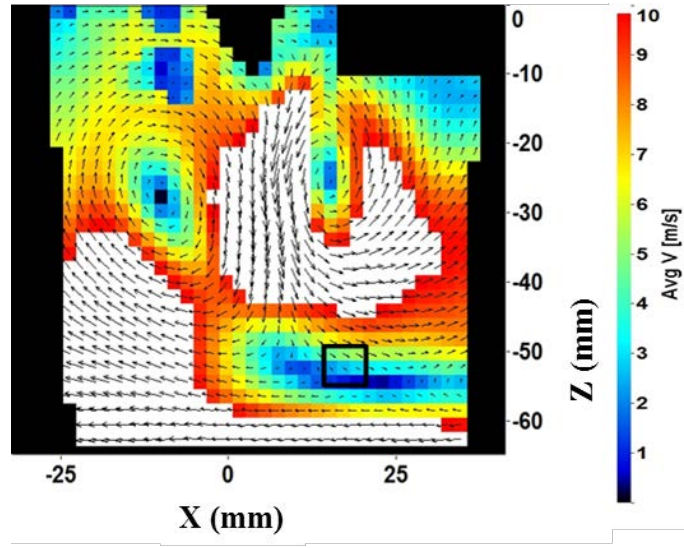

*Figure 20: Region (outlined in black) where spatial average velocity components are used differentiate between flow patterns at 100° ATDCE in Data Set #1*

Figure 21 shows the spatial average velocity components from the region illustrated in Figure 20 and the engine speed transients for Data Set #1. It can be seen that a sign change in $\bar{U}$ and $\bar{W}$ occurs between Cycles 1500 and 2300 (pink region in Figure 21), which is during the RPM perturbation between cycles 1250 and 2100. This reveals the change in the more-probable downward flow pattern of SS 1 to the more probable upward flow pattern of SS 2, which are demarcated by the green lines in Figure 21. Two possible hypotheses for the shift in flow pattern can be suggested based on the information available in this study: The RPM transient could have (a) triggered a flow instability that in turn changed the in-cylinder swirl, or (b) triggered the lateral valve oscillations that then led to a change in swirl ratio. Of course, there are other possibilities that could trigger this change; however, it is not possible to definitively assess the initiating cause of the cyclic variability using only the currently available data.

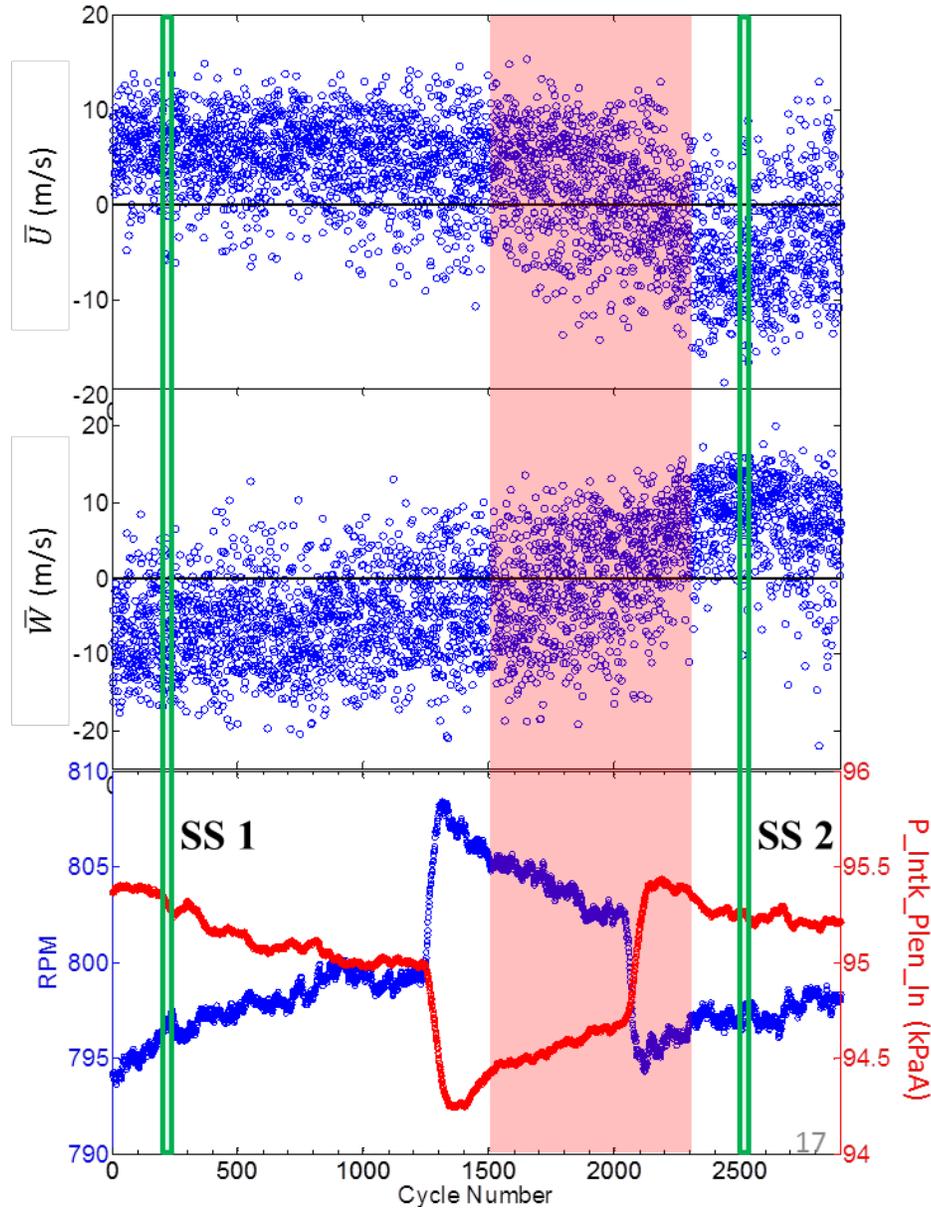

*Figure 21: Correlation between in-cylinder flow patterns at 100° ATDCE and engine speed transients in Data Set #1. Subsets 1 and 2 shown in green, engine speed transient shown in pink.*

**SUMMARY**

This paper investigates possible causes of cycle-to-cycle flow variability using both two-dimensional PIV and three-dimensional LES data. Test-to-test variation, intra-cycle variation, and cycle-to-cycle flow variations were observed in the experimental data. A comprehensive study focused on the flow observed in a plane cutting through the intake jet during mid-stroke (100° ATDCE), where a single dominant flow pattern is present in the lower right-hand corner of the field-of-view with either an upward or downward direction.

Four data samples are presented, two with an ensemble average upward flow pattern and two with a downward flow pattern. Two of the data samples were taken from two separate tests with each test having one or the other flow pattern direction on average. The other two data samples were taken from the same 2900-cycle test where one or the other flow pattern direction was dominant at different times during the test. Phase-dependent POD provided a metric to identify the existence of one or the other flow pattern. Although each data set was dominated by one of these two flow patterns, both flow patterns existed in each data set. The results demonstrate that CCV in this engine can vary from cycle to cycle, test-to-test and exist throughout the intake stroke. The appearance of one or the other of the two flow patterns in the ensemble-average velocity is sample dependent. Further, the intra-cycle PIV measurements demonstrated that the flow-pattern switching could appear in consecutive cycles.

Examination of the three-dimensional LES velocity data showed that the PIV plane was cutting through the gradient region between the intake jet and another large structure, with large out-of-plane velocity and significant variability in the volume-averaged swirl ratio. These observations suggested the azimuthal rotation of these large-scale structures. Rotation of the 2-D data sampling plane ± 10° (with respect to the cylinder symmetry axis) within the 3-D LES data demonstrated the same flow variation observed in the 2-D plane of the measured cycles. This led to the conclusion that the observed flow pattern variability might be a manifestation of large-scale flow rotation between cycles. POD analysis was used to quantify the flow pattern variability and used to assess the sensitivity of the flow variability to the incidental variations of intake port pressure, lateral valve motion and engine speed perturbations.

Cycle-to-cycle variations in the pressure at the intake port and the pressure differential between the intake port and the cylinder were compared to the cycle-to-cycle variations in the phase-dependent POD coefficients that quantified differences in flow pattern. There was no correlation between the intake port pressure and the flow pattern in the central tumble plane at 100° ATDCE. No correlation was seen between the pressure differential across the intake valve and the flow pattern at 100° ATDCE either.

An incidental lateral oscillation of the intake valve as it opened and closed was investigated as a possible cause of the observed flow CCV at 100° ATDCE. The results here are not definitive since correlation between the out-of-plane motion was not detected and the oscillations, due to a worn valve guide, could not be controlled deterministically. However, a weak correlation between the ensemble-average intake-valve position and in-cylinder flow pattern did exist for one of these four data sets presented here. It was seen that, on average, cycles where the intake valve was closer to the centerline of the cylinder, the upward velocity vectors from the bottom-right of the field-of-view were of larger magnitude. .

An approximately ±8 rpm perturbation during the 2900-cycle test was investigated as a possible trigger for the change in the flow pattern. It was found that the dominant flow pattern changed from the downward direction observed in the data sample taken before the RPM perturbation, to the upward direction observed after the perturbation. No direct correlation was observed between the engine speed and the flow pattern switching between ensemble-averaged cycles sampled prior to and after the engine speed perturbation. Further, no correlation could be associated with individual cycles.

This study does not provide a definitive relationship between cyclic variability of the in-cylinder flow and a forcing function. The switching between two flow patterns observed in this engine, couple with the incidental variations in the intake-pressure, valve oscillations, and engine speed provided an opportunity to investigate possible causes forcing the CCV. Further, it is demonstrated that POD can be used as a metric to quantify the cycle variability of the flow and provides a procedure to evaluate the correspondence between the flow variability and the engine boundary conditions. Though not definitive, the results show no obvious or overwhelming cause-and-effect relationship exists for the three sources studied here. Nonetheless, this does provide a method and guidance for future experiments in this engine, which should seek alternative sources.

**ACKNOWLEDGMENTS**

**REFERENCES**

1. Baby X, Dupont A, Ahmed A, Deslandes W, Charnay G and Michard M. A New Methodology to Analyze Cycle-to-Cycle Aerodynamic Variations. *SAE 2002-01-2837*. 2002.
2. Ghandhi JB, Herold RE, Shakal JS and Strand TE. Time-Resolved Particle Image Velocimetry Measurements in an Internal Combustion Engine. *SAE Paper 2005-01-3868*. 2005.
3. Jarvis S, Justham T, Clarke A, Garner CP, Hargrave GK and Richardson D. Time Resolved Digital PIV Measurements of Flow Field Cyclic Variation in an Optical IC Engine. *SAE Paper 2006-01-1044*. 2006.
4. Amelio M, Bova S and Bartolo CD. The Separation Between Turbulence and Mean Flow in ICE LDV Data: The Complementary Point-of-view of Different Investigation Tools. *Journal of Engineering for Gas Turbines and Power*. 2000; 122: 579-87.
5. Li Y, Zhao H, Leach B, Ma T and Ladommatos N. Characterization of an in-cylinder flow structure in a high-tumble spark ignition engine. *International Journal of Engine Research*. 2004; 5: 375-400.
6. St.Hill N, Asadamongkon P and C.Lee K. A study of turbulence and cyclic variation levels in internal combustion engine cylinder. *10th International Symposium on Application of Laser Techniques to Fluid Mechanics*. Lisbon, Portugal2000.
7. Liu D, Wang T, Jia M and Wang G. Cycle-to-cycle variation analysis of in-cylinder flow in a gasoline engine with variable valve lift. *Experiments in Fluids*. 2012; 53: 585-602.
8. Li Y, Zhao H, Peng Z and Ladommatos N. Analysis of Tumble and Swirl Motions in a Four-Valve SI Engine. *SAE paper 2011-01-3555*. 2001.
9. Elzahaby AM, Elshenawy EA and Gadallah AH. Cyclic Variability in I.C. Engines: Insights from Particle Image Velocimetry Measurements. *6th International Symposium on Diagnostics and Modeling of Combustion in Internal Combustion Engines (COMODIA 2004)*. Yokohama, Japan: Japan Society of Mechanical Engineers, 2004.
10. Voisine M, Thomas L, Boree J and Rey P. Spatio-temporal structure and cycle to cycle variations of an in-cylinder tumbling flow. *Experiments in Fluids*. 2011; 50: 1393-407.
11. Vu T-T and Guibert P. Proper orthogonal decomposition analysis for cycle-to-cycle variations of engine flow. Effect of a control device in an inlet pipe. *Experiments in Fluids*. 2012; 52: 1519-32.
12. Cosadia I, Borée J, Charnay G and Dumont P. Cyclic variations of the swirling flow in a Diesel transparent engine. *Experiments in Fluids*. 2006; 41: 115-34.
13. Enaux B, Granet V, Vermorel O, et al. LES study of cycle-to-cycle variations in a spark ignition engine. *Proceedings of the Combustion Institute*. 2011; 33: 3115-22.
14. Chen H, Reuss DL and Sick V. A practical guide for using proper orthogonal decomposition in engine research. *International Journal of Engine Research*. 2013; 14: 307-19.
15. Chen H, Reuss DL and Sick V. On the use and interpretation of proper orthogonal decomposition of in-cylinder engine flows. *Meas Sci Technol*. 2012; 23: 085302.
16. Haworth DC. Large-eddy simulation of in-cylinder flows. *Oil & Gas Science and Technology*. 1999; 54: 175-85.
17. Haworth DC. A review of turbulent combustion modeling for multidimensional in-cylinder CFD. *SAE Paper 2005-01-0993*. 2005.
18. Haworth D and Jansen K. Large-eddy simulation on unstructured deforming meshes: Towards reciprocating IC engines. *Computers and Fluids*. 2000; 29: 493-524.
19. ElTahry SH and Haworth D. Directions in turbulence modeling for in-cylinder flows in reciprocating engines. *AIAA Journal of Propulsion and Power*. 1992; 8: 1040-8.

20. Celik IB, Amin E, Smith J, Yavuz I and Gel A. Towards large eddy simulation using the KIVA code. *11th International Multidimensional Engine Modeling User's Group Meeting*. Detroit, Michigan1998, p. February 1998.
21. Celik IB, Yavuz I, Smirnov A, Smith J, Amin E and Gel A. Prediction of in-cylinder turbulence for IC engines. *Combustion Science and Technology*. 2000; 153: 339-68.
22. Naitoh K, Kaneko Y and Kayuza I. SI-Engien Design Concept for Reducing Cyclic Variations. *SAE PAper 2005-01-0992*. 2005.
23. Richard S, Colin O, Vermorel O, Benkenida A, Angelberger C and Veynante D. Towards large eddy simulation of combustion in spark ignition engines. *Proceedings of the Combustion Institute*. 2007; 31: 3059-66.
24. Abraham PS, Liu K, Haworth DC, Reuss Dl and Sick V. Evaluating LES and PIV with Phase-Invariant POD. *Oil and Gas Science and Technology*. 2014; 69: 41 - 59.
25. Schiffmann P, Gupta S, Reuss D, Sick V, Yang X and Kuo T-W. TCC-III Engine Benchmark for Large Eddy Simulation of IC Engine Flows: submitted. *Oil & Gas Science and Technology*. 2014: submitted.
26. Sick V, Reuss D, Rutland C, et al. A Common Engine Platform for Engine LES Development and Validation. In: Angelberger C, (ed.). *LES4ICE*. Rueil-Malmaison, France: IFP Energies Nouvelles, 2010.
27. Yang X, Gupta S, Kuo T-W and Gopalakrishnan V. RANS and LES of IC Engine Flows - A comparative study. *ASME Internal Combustion Engine Fall Technical Conference*. Dearborn, Michigan, USA2013, p. ICEF2013-19043.
28. Kuo T-W, Yang X, Gopalakrishnan V and Chen Z. Large Eddy Simulation (LES) for IC Engine Flows. *Oil & Gas Science and Technology*. 2014; 69: 61 - 81.
29. Lumley JL. *Engines, An Introduction*. Cambridge University Press, 1999.
30. Pera C and Angelberger C. Large Eddy Simulation of a Motored Single-Cylinder Engine Using System Simulation to Validate Boundary Conditions. SAE2011.
31. Arcoumanis C, Bicen AF, Vafidis C and Whitelaw JH. Three-Dimensional Flow Field in Four-Stroke Model Engines. *SAE Paper 841360*. 1984.
32. Khalighi B, ElTahry S, Haworth D and Huebler M. Computation and Measurement of Flow and Combustion in a Four-Valve Engine with Intake Variation. *SAE Paper 950287*. 1995.
33. Kim M and Ohm I. The effect of intake valve angle on in-cylinder flow during intake and compression process. *SAE Paper 2007-01-4045*. 2007.
34. Reuss DL, Kuo T, Khalighi B, Haworth D and Rosalik M. Particle Image Velocimetry Measurements in a High-Swirl Engine Used for Evaluation of Computational Fluid Dynamics Calculations. *SAE Paper 952381*. 1995.
35. Reuss DL and Rosalik ME. PIV Measurements during Combustion in a Reciprocating Internal Combustion Engine. *9th International Symposium on Applications of Laser Technologies in Fluid Mechanics*. Lisbon, Portugal: Springer Verlag, 1998.
36. Reuss DL. Cyclic Variability of Large-Scale Turbulent Structures in Directed and Undirected IC Engine Flows. *SAE Paper 2000-01-0246*. 2000.
37. Abraham PS, Reuss DL and Sick V. High-speed particle image velocimetry study of in-cylinder flows with improved dynamic range. *SAE International*. 2013, p. SAE Paper 2013-01-0542.
38. CONVERGE™: A Three-Dimensional Computational Fluid Dynamics Program for Transient Flows with Complex Geometries. *Convergent Science Inc*. 2009.
39. Yoshizawa A and Horiuti K. A statistically-Derived Subgrid-Scale Kinetic Energy Model for the Large-Eddy Simulation of Turbulent Flows. *J Phys Soc Jpn*. 1985; 54: 2834-9.

40. Menon S, K. YP and W. KW. Effect of Subgrid Models on the Computed Interscale Energy Transfer in Isotropic Turbulence. *Computers and Fluids*. 1996; 25: 165-80.
41. Ghosal S, S. LT, P. M and K. A. A dynamic localization model for large-eddy simulation of turbulent flows. *J Fluid Mech*. 1995; 286: 229-55.
42. Pomraning E, Rutland and J. C. A Dynamic One-Equation Non-Viscosity LES Model. *AIAA Journal*. 2002; 40: 689-701.
43. Werner H and Wengle H. Large-Eddy Simulation of Turbulent Flow Over and Around a Cube in a Plate Channel. *Turbulent Shear Flows*. 1993; 8: 155-68.
44. Issa RI. Solution of the implicitly discretised fluid flow equations by operator-splitting. *Journal of Computational Physics*. 1986; 62: 40-65.
45. Fogleman M, Lumley J, Rempfer D and Haworth D. Application of the Proper Orthogonal Decomposition to Datasets of Internal Combustion Engine Flows. *Journal of Turbulence*. 2004; 5.
46. Chen H, Reuss DL and Sick V. Analysis of misfires in a direct injection engine using Proper-Orthogonal Decomposition. *Experiments in Fluids*. 2011; 51: 1139-51.
47. Liu K and Haworth D. Development and Assessment of POD for Analysis of Turbulent Flow in Piston Engines. *SAE Paper 2011-01-0830*. 2011.